\documentclass[twocolumn]{pasj00}
\usepackage[dvips]{color}
\usepackage{multicol}

\begin{document}
\SetRunningHead{N. Narita et al.}{The Possible Spin-Orbit
Misalignment of HD~17156b}
\Received{2007/12/16}
\Accepted{2008/01/30}

\title{A Possible Spin-Orbit Misalignment\\
in the Transiting Eccentric Planet HD~17156b$^{*}$}


\author{
Norio \textsc{Narita},\altaffilmark{1}$^{**}$
Bun'ei \textsc{Sato},\altaffilmark{2}
Osamu \textsc{Ohshima},\altaffilmark{3} and
Joshua N.\ \textsc{Winn}\altaffilmark{4}
}

\altaffiltext{1}{
Department of Physics, School of Science, The University of Tokyo,
7-3-1 Hongo, Bunkyo-ku, Tokyo 113--0033, Japan
}
\email{narita@utap.phys.s.u-tokyo.ac.jp}

\altaffiltext{2}{
Global Edge Institute, Tokyo Institute of Technology,
2-12-1 Ookayama, Meguro, Tokyo 152-8550, Japan
}

\altaffiltext{3}{
Mizushima Technical High School,
1230 Nishiachi, Kurashiki, Okayama 710-0807, Japan
}

\altaffiltext{4}{
Department of Physics, and Kavli Institute for Astrophysics
and Space Research,\\
Massachusetts Institute of Technology, Cambridge, MA 02139, USA
}

\KeyWords{
stars: planetary systems: individual (HD~17156) ---
stars: rotation --- 
techniques: photometric ---
techniques: radial velocities --- 
techniques: spectroscopic}

\maketitle

\begin{abstract}
  We present simultaneous photometric and spectroscopic observations
  of HD~17156b spanning a transit on UT 2007 November 12. This system is
  of special interest because of its 21-day period (unusually long for
  a transiting planet) and its high orbital eccentricity of 0.67. By
  modeling the Rossiter-McLaughlin effect, we find the angle between
  the sky projections of the orbital axis and the stellar rotation
  axis to be $62^{\circ} \pm 25^{\circ}$.
  Such a large spin-orbit misalignment, as
  well as the large eccentricity, could be explained as the relic of a
  previous gravitational interaction with other planets.
\end{abstract}
\footnotetext[*]{Based in part on data collected at the Okayama
Astrophysical Observatory,
which is operated by the National Astronomical Observatory of Japan.}
\footnotetext[**]{JSPS Fellow.}

\section{Introduction}

More than 250 exoplanets are now known, and their
orbital characteristics are remarkably diverse.
In particular, the surprising discoveries of
close-in giant planets and planets on highly eccentric orbits
have led to interesting revisions of planet formation theory.
It is generally believed that the close-in giant planets originally
formed at larger orbital distances (beyond the ``snow line'') and
migrated inward during the planet formation epoch
\citep{1996Natur.380..606L}.  One possible migration mechanism is
tidal interaction with the protoplanetary gas disk, after the planet
opens up a gap in the disk (Type II migration; see, e.g.,
\cite{1985prpl.conf..981L}), which is appealing because it can account
for the observed period distribution (see, e.g.,
\cite{2004ApJ...616..567I}).  However, some planets have large
eccentricities despite their close-in orbits, such as HAT-P-2b
\citep{2007ApJ...670..826B}, GJ~436b \citep{2004ApJ...617..580B}, and
HD~17156b \citep{2007ApJ...669.1336F}.  Since disk-planet interaction
would not excite a planet's eccentricity to such a level
\citep{2004ApJ...606L..77S}, the scenario does not provide an obvious
explanation for these systems.  On the other hand, some other
migration models would naturally produce eccentric orbits, namely,
planet-planet gravitational scattering
(\cite{1996Sci...274..954R, 2002Icar..156..570M}; Nagasawa et al.~2008).
In these models, planets would initially obtain very large
eccentricities and small periastron distances through the
gravitational interactions, and subsequently evolve into
shorter-period and more circular orbits through tidal dissipation.

Assuming that planets are initially formed on circular orbits,
the observation of highly eccentric orbits is already an indication
of previous gravitational interactions. What other type of evidence
might establish the case for such interactions?
Recently, the alignment angle of the stellar spin axis and the planetary
orbital axis (the spin-orbit alignment angle) has become recognized to be
a promising diagnostic.
This is because disk-planet interaction would probably
maintain the original spin-orbit alignment,
while planet-planet scattering
would cause significant misalignments in a nontrivial fraction
of cases (e.g., \cite{2007astro.ph..3166C}; Nagasawa et al. 2008).
Because the damping time scale of the spin-orbit alignment angle
is expected to be longer (by a few orders of magnitude) than
that of eccentricity \citep{1981A&A....99..126H, 2006ApJ...653L..69W},
one might regard the spin-orbit angle
as a fossil record of planetary migration.

It is possible to measure the sky-projection of the
spin-orbit alignment angle, $\lambda$, for
a transiting exoplanet, by making use of
the Rossiter-McLaughlin effect (hereafter the RM effect:
\cite{1924ApJ....60...15R}, \cite{1924ApJ....60...22M}).
The RM effect is the radial velocity (RV) anomaly
caused by the partial occultation of the
rotating stellar surface (see \cite{2005ApJ...622.1118O},
\cite{2007ApJ...655..550G} for theoretical descriptions), and
several observers have reported the detections of the RM effect and
measured $\lambda$ in transiting planetary systems
\citep{2000A&A...359L..13Q, 2005ApJ...631.1215W, 2006ApJ...653L..69W,
2007PASJ...59..763N, 2007ApJ...665L.167W, 2007arXiv0707.0679L,
2007ApJ...667..549W}.
However, the reported angles for the spin-orbit alignment were either
small or consistent with zero, even for the highly eccentric case of
HAT-P-2b \citep{2007ApJ...665L.167W, 2007arXiv0707.0679L}.

In this Letter, we present a photometric and spectroscopic study of a
recent transit of HD~17156b \citep{2007arXiv0710.0898B}.  This planet
has a large orbital eccentricity and a small periastron distance, and
is therefore a promising candidate for a spin-orbit misalignment.
The system has also attracted considerable interest because it has by
far the longest period (21 days) of any known transiting exoplanet.
We describe our observations in Section 2 and our results in Section 3.
The final section is a summary and a discussion of the results.

\section{Observations and Data Reduction}

We observed a transit of HD~17156b on UT 2007 November 12, both
spectroscopically and photometrically, with telescopes in the Okayama
prefecture of Japan. The transit was predicted to occur in the later
part of that night, according to ephemerides kindly provided by
D.~Charbonneau (2007, private communication).  The photometric transit
was observed with a 20~cm Meade telescope at Kurashiki-shi.  The
spectroscopic transit was observed with the 188~cm telescope at the
Okayama Astrophysical Observatory (OAO).  We observed HD~17156 for a
5~hr period spanning the transit time, with both telescopes, through
air masses ranging from 1.3 to 1.6.

\subsection{Photometry}

The photometric observations were conducted at Kurashiki-shi
(E\timeform{133D40'15"}, N\timeform{34D32'27"}, h=7m) using a 20~cm
Meade LX200R-20 $f/10$ telescope.  A cooled (253~K) ST-9XE CCD camera
provides a $17.6 \times 17.6$ arcmin$^2$ field of view with the pixel
scale of 2.06 arcsec.  We defocused the stellar images in order to
average over pixel-to-pixel sensitivity variations, and to draw out
the exposure time and thereby increase the duty cycle.  The exposure
time was 60~s and the readout time was 1~s.  We observed through a Rc
band filter covering the range 5720--6780~\AA.

We used the Astronomical Image Processing for Windows software
(AIP4Win Ver.2.1.10) for the subtraction of dark current,
flat-fielding, and aperture photometry.  We determined the apparent
magnitudes of HD~17156 ($V=8.17$) and two comparison stars: HD~16906
($V=8.28$) and BD+71~168 ($V=9.57$) using an aperture radius of 20
pixels.  The typical FWHM of each star ranged from 18 to 21 pixels
(from 37 to 43 arcsec).  We estimated the sky background level with an
annulus from 27 to 32 pixels in radius centered on each star, and
subtracted the estimated sky contribution from the aperture flux.
Then we computed the differential magnitude between HD~17156 and the
ensemble average of the comparison stars.  After these steps, we
clearly detected the transit event.  We also found small linear trend
in the out-of-transit (OOT) data; we removed the trend by fitting the
OOT data with a linear function of time.

\begin{figure}[thb]
 \begin{center}
  \FigureFile(45mm,45mm){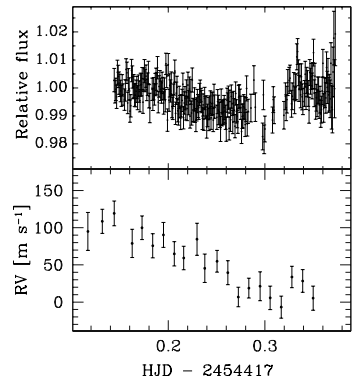}
 \end{center}
  \caption{Top: Photometric light curve observed with the
  20~cm telescope (251 samples) on UT 2007 November 12.
  The error-bars are scaled to satisfy $\chi^2/\nu_{dof} = 1.0$.
  Bottom: Radial velocity samples on that night computed from the
  OAO/HIDES spectra.
  The values and uncertainties are presented in Table 1.
  \label{ourdata}}
\end{figure}

We estimated photometric errors of our samples as follows.  We first
calculated the Poisson noise of each sample and found that the scatter
for the OOT data is systematically larger than the Poisson error.  In
order to account for the systematic errors in the photometry, we
scaled the Poisson estimates of the errors to satisfy
$\chi^2/\nu_{dof} = 1.0$ for the OOT dataset.  Next, we also
considered the time-correlated noise (the so-called ``red noise'', see
e.g., \cite{2006MNRAS.373..231P}).  We compared the standard deviation
of each OOT sample ($\sigma_1$) with the standard deviation of the OOT
data after averaging each $N$ points ($\sigma_N$).  We found that
$\sigma_N < \sigma_1 / \sqrt{N}$ for values of $N$ ranging from 10 to
60, suggesting that the effects of correlated noise are small in this
case.  Thus we did not modify the error bars further.  The typical
photometric error is about 4 mmag.  Our light curve is shown in the
top panel of Figure~\ref{ourdata}.

\subsection{Spectroscopy and Radial Velocity}

We used the High Dispersion Echelle Spectrograph (HIDES;
\cite{1999PYunO....S..77I}) on the 188~cm telescope at OAO.  We
employed the standard setup for RV measurements with the Iodine
absorption cell (covering the wavelength range 5000--6200~\AA, where
the iodine absorption lines are abundant).  The slit width of 280
$\mu$m ($1\farcs06$) yielded a spectral resolution of $\sim$45000, and
the seeing was between $1\farcs5$ and $2\farcs0$.  For the transit
event of UT~2007~November~12, we used 15~min exposures with a binnning
of 4 (spatial) $\times$ 1 (dispersion).  We also obtained a few
spectra on UT~2007 November 14, 17, and 18, outside of transits,
in order to refine the spectroscopic orbit.  For these OOT spectra we
used 30~min exposures and 1 $\times$ 1 binning.
The typical signal-to-noise ratio (SNR) was 50-80 pixel$^{-1}$.
We processed the frames with standard IRAF\footnote{The Image
Reduction and Analysis Facility (IRAF) is distributed by the U.S.\
National Optical Astronomy Observatories, which are operated by the
Association of Universities for Research in Astronomy, Inc., under
cooperative agreement with the National Science Foundation.}
procedures and extracted one-dimensional spectra.
We then calculated relative RV variations by the algorithm
following \citet{2002PASJ...54..873S}.
We estimated internal errors of the radial velocities from the scatter of
the RV solutions for 4-5~\AA~segments of the spectra.
The typical errors are 10-20~m~s$^{-1}$.
Our radial velocities are presented in Table~\ref{rvsummary}, and
plotted in the bottom panel of Figure~\ref{ourdata}.

\begin{table}[t]
\caption{Radial velocities obtained with the OAO/HIDES.}
\begin{center}
\begin{tabular}{lcc}
\hline
Time [HJD]  & Value [m~s$^{-1}$] & Error [m~s$^{-1}$]\\
\hline
2454417.11678 &	94.98 	& 25.53 \\
2454417.13164 &	108.60 	& 16.24 \\
2454417.14398 &	119.28 	& 16.71 \\
2454417.16256 &	78.87 	& 19.01 \\
2454417.17275 &	99.93 	& 15.94 \\
2454417.18389 &	75.77 	& 16.24 \\
2454417.19495 &	90.34 	& 16.74 \\
2454417.20621 &	64.90 	& 16.37 \\
2454417.21594 &	59.26 	& 15.77 \\
2454417.22968 &	84.55 	& 21.55 \\
2454417.23773 &	45.40 	& 19.12 \\
2454417.25048 &	54.91 	& 14.57 \\
2454417.26162 &	39.53 	& 16.10 \\
2454417.27249 &	6.80 	& 13.14 \\
2454417.28340 &	18.77 	& 13.29 \\
2454417.29502 &	21.32 	& 19.48 \\
2454417.30549 &	5.71 	& 15.86 \\
2454417.31692 &	-6.74 	& 14.82 \\
2454417.32796 &	33.70 	& 14.45 \\
2454417.33911 &	28.36 	& 15.27 \\
2454417.34993 &	5.32 	& 16.25 \\
2454419.08778 & -140.28 & 10.18 \\
2454421.92713 &	36.65 	& 10.99 \\
2454422.17592 &	66.31 	& 18.30 \\
2454423.13311 &	88.78 	& 12.96 \\
\hline
\end{tabular}
\label{rvsummary}
\end{center}
\end{table}

\section{Analyses and Results}

As described in the previous section, we have obtained 25 RV
samples and 251 Rc band photometric samples covering the transit.
In addition, in order to determine an optimal solution of orbital
parameters for HD~17156, we incorporate previously published radial
velocity data taken at the Subaru telescope and the Keck~I telescope
(9 from Subaru and 24 from Keck: \cite{2007ApJ...669.1336F}).

We employ the analytic formulas for light curves and radial velocities
including the RM effect given in \citet{2005ApJ...622.1118O} and
\citet{2006astro.ph.11466O} (hereafter the OTS formulae) in order to
model the observed data.  We also adopt the correction for $V \sin
I_s$ in the OTS formulas by modifying $V \sin I_s \textrm{(OTS)} = V
\sin I_s \textrm{(real)} * 1.1$; see \cite{2005ApJ...631.1215W} and
\cite{2007PASJ...59..763N} for details.  (We note that this correction
does not affect the results for $\lambda$, since in this case
$\lambda$ and $V\sin I_s$ are nearly uncorrelated parameters.)
We adopt the stellar mass $M_s = 1.2$ [$M_{\odot}$] and
the stellar radius $R_s = 1.47$ [$R_{\odot}$]
\citep{2007ApJ...669.1336F}, and the quadratic limb-darkening coefficient
$u_{1,r} = 0.31$ and $u_{2,r} = 0.35$ for the Rc band photometry, and
$u_{1,g} = 0.49$ and $u_{2,g} = 0.28$ for the spectroscopic band,
based on \citet{2004A&A...428.1001C}.
The adopted parameters are summarized in Table~\ref{parameter}.

\begin{table}[t]
\caption{Star and planet parameters.}
\begin{tabular}{l|cc}
\hline
Parameter & Value & Source \\
\hline
$M_s$ [$M_{\odot}$] 
& $1.2\,\,^{\rm{a}}$  & \cite{2007ApJ...669.1336F} \\
$R_s$ [$R_{\odot}$]
& $1.47\,\,^{\rm{a}}$ & \cite{2007ApJ...669.1336F} \\
$u_{1,r}$ 
& $0.31\,\,^{\rm{a}}$ & \cite{2004A&A...428.1001C} \\
$u_{2,r}$ 
& $0.35\,\,^{\rm{a}}$ & \cite{2004A&A...428.1001C} \\
$u_{1,g}$ 
& $0.49\,\,^{\rm{a}}$ & \cite{2004A&A...428.1001C} \\
$u_{2,g}$ 
& $0.28\,\,^{\rm{a}}$ & \cite{2004A&A...428.1001C} \\
$a$ [A.U.]
& $0.15\,\,^{\rm{a}}$ & \cite{2007ApJ...669.1336F} \\
$P$ [days]
& $21.2162 \pm 0.0036$ & This work.$^{\rm{b}}$ \\
$M_p$ [$M_{Jup}$]
& $3.13 \pm 0.21$ & This work.$^{\rm{c}}$ \\
$R_p$ [$R_{Jup}$]
& $1.21 \pm 0.12$ & This work.$^{\rm{c}}$ \\
\hline
\multicolumn{3}{l}{\hbox to 0pt{\parbox{80mm}{\footnotesize
\footnotemark[a]:Adopted.\\
\footnotemark[b]:Determined in this work
thanks to D. Charbonneau 2007, private communication.\\
\footnotemark[c]:Determined from the values and errors in this letter
and the error of $M_s$ and $R_s$ presented in \cite{2007ApJ...669.1336F}.
\\
}\hss}}
\end{tabular}
\label{parameter}
\end{table}

Our model has 11 free parameters in total.
Seven parameters for the HD~17156 system include the RV
amplitude $K$, the eccentricity $e$, the longitude of periastron $\omega$,
the sky-projected stellar rotational velocity $V \sin I_s$,
the sky-projected angle between the stellar spin and the planetary
orbit axes $\lambda$, the ratio of star-planet radii $R_p/R_s$,
and the orbital inclination $i$.
We also add three parameters
for velocity offsets to the respective RV dataset $v_{1}$
(for our template spectrum), $v_{2}$ (for the Subaru data)
and $v_{3}$ (for the Keck data), and one parameter for the time of
mid-transit $T_c$ on UT 2007 November 12.

As a first step, we determined $T_c$ using our photometric data only,
and used this result to refine the estimate of the orbital period $P$,
with reference to a previous transit epoch (D.~Charbonneau 2007,
private communication). We found $T_c = 2454417.2645 \pm 0.0022$
[HJD], and thereby $P = 21.2162 \pm 0.0036$ [days].  We adopt $P =
21.2162$ for the subsequent analysis (the uncertainty is negligible
for our purpose).

Next, we determined the orbit of HD~17156 by simultaneous
fitting of the photometric data and the RV data.
Our $\chi^2$ fitting statistic is
\begin{eqnarray}
\chi^2 &=& \sum_{i=1}^{N_{rv}=58} \left[ \frac{v_{i,obs}-v_{i,calc}}
{\sigma_{i}} \right]^2 + \sum_{j=1}^{N_{f}=251} \left[ \frac{f_{j,obs}
-f_{j,calc}}{\sigma_{j}} \right]^2 \nonumber\\
&+& \left[ \frac{V \sin I_s - 2.6}{0.5} \right]^2,
\end{eqnarray}
where $v_{calc}$ and $f_{calc}$ represent the values calculated by the
OTS formulae with the above parameters.  The last term is \textit{a
  priori} constraint on $V \sin I_s$, which enforces the spectroscopic
determination by \citet{2007ApJ...669.1336F}.  We found optimal
orbital parameters by minimizing the $\chi^2$ statistic using the
AMOEBA algorithm \citep{1992nrca.book.....P}, and estimated confidence
levels by $\Delta \chi^2$ as the parameters were stepped
away from the optimal values.  (We also estimated the parameter
uncertainties with bootstrap and Markov Chain Monte Carlo methods, as
described below.)

\begin{figure}[t]
 \begin{center}
  \FigureFile(80mm,70mm){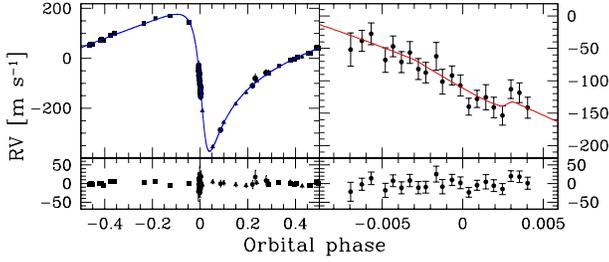}
 \end{center}
  \caption{
  Radial velocities of HD~17156 as a function of orbital phase, and the
  best-fitting model
  with the \textit{a priori} constraint on $V \sin I_s$.
  Three symbols represent the different dataset respectively
  (circle: OAO, triangle: Subaru, square: Keck).
  Left panel: The entire orbit.
  Right panel: A zoom of transit phase.
  Bottom panels: Residuals from the best-fit curve.
  \label{rvfit}}
\end{figure}

To account for possible systematic errors in the RV measurements (from
photospheric jitter or instrumental sources), we used the following
procedure. First, we determined the optimal parameter set for the OOT
data using the nominal RV errors, and calculated the $\chi^2$ from
each RV dataset (OAO, Subaru, and Keck).  We then computed
$\chi^2/\nu_{dof}$, where $\nu_{dof}$ is the number of each dataset
minus 4, which is the number of related parameters for the RV data
($K$, $e$, $\omega$, and each RV offset).  Next, when
$\chi^2/\nu_{dof} > 1$, we inflated the radial velocity errors to
satisfy $\chi^2/\nu_{dof} = 1.0$.  After these steps, we re-calculated
optimal parameters and uncertainties using all datasets.  We also
computed the results for the $\chi^2$ statistic without the \textit{a
  priori} constraint on $V\sin I_s$, to support our claim that the
application of this constraint does not affect the results for
$\lambda$.  The resultant parameters are presented in
Table~\ref{result}\,\footnote{Note that we present only $V \sin I_s$
  and $\lambda$ as the results without the \textit{a priori}
  constraint in the table},
and the RV curve with the best-fit model is shown in
Fig~\ref{rvfit}.  Consequently, our result for the key parameter
$\lambda$ was $\lambda = 62^{\circ} \pm 25^{\circ}$ ($60^{\circ} \pm
21^{\circ}$) with the reduced $\chi^2$ of 0.97 (0.96).  The numbers in
parentheses refer to the case without the \textit{a priori}
constraint.
The value of $\lambda$ is fairly large, and is inconsistent with zero
at the $\sim$2.5$\sigma$ level,
indicating a possibly large spin-orbit misalignment.

\begin{table}[t]
\caption{Fitted values and uncertainties$^{\rm{a}}$ of the parameters.}
\begin{center}
\begin{tabular}{l|cc}
\hline
Parameter & Value & Uncertainty \\
\hline
$K$ [m s$^{-1}$] 
& $274.7$ & $\pm 3.5$ \\
$e$ 
& $0.675$ & $\pm 0.005$ \\
$\omega$ [$^{\circ}$]
& $122.0$ & $\pm 0.6$ \\
$R_p/R_s$ 
& $0.0846$ & $\pm 0.0026$ \\
$i$ [$^{\circ}$] 
& $85.65$ & $\pm 0.29$ \\
$v_1$ [m s$^{-1}$] 
& $146.7$ & $\pm 3.9$ \\
$v_2$ [m s$^{-1}$] 
& $92.9$ & $\pm 2.6$ \\
$v_3$ [m s$^{-1}$] 
& $93.4$ & $\pm 1.2$ \\
$T_c - 2454417$ [HJD] 
& $0.2645$ & $\pm 0.0021$ \\
$V \sin I_s \,\,^{\rm{b}}$ [km s$^{-1}$]
& $2.8$  & $\pm 0.5$ \\
$\lambda\,\,^{\rm{b}}$ [$^{\circ}$]\,\, ($\Delta \chi^2$)
& $62$  & $\pm 25$ \\
$\lambda\,\,^{\rm{b}}$ [$^{\circ}$]\,\, (bootstrap)
& $67$  & $\pm 26$ \\
$\lambda\,\,^{\rm{b}}$ [$^{\circ}$]\,\, (MCMC)
& $65$  & $\pm 25$ \\
\hline
$V \sin I_s \,\,^{\rm{c}}$ [km s$^{-1}$]
& $4.7$  & $\pm 1.5$ \\
$\lambda\,\,^{\rm{c}}$ [$^{\circ}$] 
& $60$  & $\pm 21$ \\
\hline
\multicolumn{3}{l}{\hbox to 0pt{\parbox{80mm}{\footnotesize
\footnotemark[a]:Computed by $\Delta \chi^2 = 1.00$.\\
\footnotemark[b]:With the \textit{a priori} constraint on $V \sin I_s$.\\
\footnotemark[c]:Without the constraint.\\
}\hss}}
\end{tabular}
\label{result}
\end{center}
\end{table}

Since the statistical significance of the result is modest, we
checked on the calculation of the confidence levels in two different
ways: a bootstrap analysis, and a Markov Chain Monte Carlo algorithm.
For the bootstrap analysis, we first computed the residuals from the
best-fitting model for UT 2007 November 12 data.  Then we scrambled
the residuals with their error-bars in a random manner, and created a
new RV dataset by adding the residuals back to the best-fitting
velocities.  In this way, we created 1000 fake data sets, and we
calculated the optimal parameters in each case, using the same method
that was applied to the actual data.  The mean values and standard
deviations of the 1000 results were
$\lambda = 67^{\circ} \pm 26^{\circ}$, which are in excellent
agreement with the $\chi^2$ analysis.  For the Markov Chain Monte
Carlo algorithm, we used a variant of the code that has been employed
by Holman et al.~(2006) and Winn et al.~(2006) for the Transit Light
Curve project (see those papers for details, or Tegmark et al.~2004
for an introduction to the method). The resulting {\it a posteriori}
probability distribution for $\lambda$ was nearly Gaussian in shape,
with mean $65\arcdeg$ and standard deviation $25\arcdeg$.

\section{Summary and Discussion}

In this Letter, we have presented the results of simultaneous
spectroscopy and photometry of a transit of HD~17156b,
which was reported to have an eccentric orbit.
(The high eccentricity is indeed confirmed by our radial velocity data
around and after the transit phase.)
We have measured the sky-projected spin-orbit alignment angle $\lambda$
by modeling the Rossiter-McLaughlin effect, and found
$\lambda = 62^{\circ} \pm 25^{\circ}$.
Although the statistical significance is modest,
this result suggests that HD~17156b has
a large spin-orbit angle.

One may wonder whether the values $e=0.67$ and $\lambda=62\arcdeg$
are physically possible and consistent with theoretical predictions.
Considering that the periastron distance $q$ for HD~17156b is
$q = a (1-e) \sim 0.05$ AU, where $a$ is the semimajor axis
and $e$ is the eccentricity, and assuming a planetary tidal quality
factor of $Q_p \sim 10^5$, then the
damping timescale for the eccentricity due to dynamic tides
in the planet is typically longer than $10$ [Gyr]
(\cite{1996ApJ...470.1187R, 2004MNRAS.347..437I}).
Thus, tidal circularization is expected to be incomplete
at its stellar age ($5.7^{+1.3}_{-1.9}$ [Gyr]:
\cite{2007ApJ...669.1336F}).
There is another possibility that the planet
initially obtained larger eccentricity and more close-in periastron
distance (say, $e > 0.99$ and $q \sim 0.04$ [A.U.]) and evolved
to the current orbit (Nagasawa et al. 2008).
In this case, the planet would be still
in the evolutionary track toward a hot Jupiter.

On the other hand, the damping of the spin-orbit angle is
mainly caused by tidal dissipation inside the host star,
rather than the planet.
Assuming a stellar tidal quality factor of
$Q_s \sim 10^6$ as a typical value, the damping timescale of the
spin-orbit alignment is estimated
to be longer than $\sim 10^{12}$ [yr] (see e.g., Eq. (15) in
\cite{2005ApJ...631.1215W}).
We note that the tidal quality factors for the star and the planet are
still unknown and the damping timescales estimated above remain fairly
uncertain.
In particular, it is possible that the coplanarization
timescale is shorter than we have calculated,
if only a thin outer convective layer is reoriented.
However, our simple considerations show that it is at least
physically plausible for the planet to have maintained
a large eccentricity and a large spin-orbit angle
since the planet formation epoch.
Moreover, very recently several theoretical simulations predicted
that certain degree of planetary systems migrated through the
planet-planet interactions would have
a large eccentricity and also a large spin-orbit misalignment
(\cite{2007astro.ph..3166C}; Nagasawa et al. 2008).
Our results are thus consistent in principle with
the predictions from these recent theoretical works.

Finally, we note that this planet is the first plausible candidate to
have a large spin-orbit misalignment.
Further RV measurements are highly desired
to bolster the precision in the determination of $\lambda$.
Such observations are perfectly feasible with a larger telescope than
the 1.9~m telescope we employed; the main difficulty is
the rarity of follow-up opportunities from a given observatory,
due to the long orbital period.
Given the present results, we believe this system and its transits are
worthy of further photometric and spectroscopic scrutiny.

\medskip

We are very grateful to Yuuki Moritani, Akira Imada, Daisaku Nogami,
and members of Okayama Planet Search Project for kindly exchanging
OAO/HIDES observing time.
We also thank David Charbonneau for providing us
the previous transit time, and Shigeru Ida and Yasushi Suto for
helpful discussions.
We appreciate the careful reading and useful comments
by the referee, Fred Rasio.
N.N. is supported by a Japan Society for Promotion of Science (JSPS)
Fellowships for Research (DC2: 18-10690), and B.S. is supported by
Grant-in-Aid for Young Scientists (B) No. 17740106.



\end{document}